\documentstyle[epsf,seceq,onecolumn]{jpsj_sissa}

\newcommand{\smint}{\!\int\!}
\newcommand{\smintq}{\smint d^2{\mib q}}
\newcommand{\smintk}{\smint d^2{\mib k}}
\newcommand{\zxi}{\xi^{{}^{(0)}}}
\newcommand{\zgamma}{\gamma^{\!{}^{(0)}}}
\newcommand{\adagger}{a^{\!{}^\dagger}}
\newcommand{\dddagger}{d^{\!{}^\dagger}}
\newcommand{\sech}{{\rm sech}}
\newcommand{\zT}{T^{{}^{(0)}}}
\newcommand{\sphi}{{\mib \phi}_\sigma}
\newcommand{\dphi}{\phi_d}
\newcommand{\bdphi}{\bar{\phi_d}}
\newcommand{\sGamma}{\Gamma_\sigma}
\newcommand{\dGamma}{\Gamma_d}
\newcommand{\schi}{\chi_\sigma}
\newcommand{\dchi}{\chi_d}
\newcommand{\uss}{u_{\sigma\sigma}}
\newcommand{\usd}{u_{\sigma d}}
\newcommand{\udd}{u_{dd}}

\def\runtitle{$d_{x^2-y^2}$ Wave Pairing Fluctuations and Pseudo Spin Gap
in Two-Dimensional Electron Systems}
\def\runauthor{Shigeki {\sc Onoda} and Masatoshi {\sc Imada}}

\title{$d_{x^2-y^2}$ Wave Pairing Fluctuations and Pseudo Spin Gap
in Two-Dimensional Electron Systems\footnote{Submitted to J. Phys. Soc. Jpn.}}
\author{Shigeki {\sc Onoda} and Masatoshi {\sc Imada}}
\inst{ Institute for Solid State Physics, University of Tokyo, Roppongi,
Minato-ku, Tokyo 106-8666}
\recdate{February 15, 1999}

\abst{Pseudogap phenomena of high-$T_c$ cuprates are examined.
In terms of AFM (antiferromagnetic) and $d$SC ($d_{x^2-y^2}$-wave
superconducting) auxiliary fields introduced to integrate out the
fermions, the effective action for 2D electron systems with AFM and
$d$SC fluctuations is considered.
By the self-consistent renormalization (SCR),
the NMR relaxation rate $T_1^{-1}$, the spin correlation length $\xi_\sigma$ 
and the pairing correlation length $\xi_d$ are calculated.
From this calculation, a mechanism of the pseudogap formation emerges
as the region of dominant $d$-wave short-range order (SRO) over
AFM-SRO.
When damping for the AFM fluctuation strongly depends on the $d$SC
correlation length through the formation of precursor singlets around
$(\pi,0)$ and $(0,\pi)$ points in the momentum space,
the pseudogap appears in a region of the normal state characterized by decreasing $1/T_1T$ and
increasing AFM correlation length with decrease in temperature.
This reproduces a characteristic feature of the pseudogap phenomena in
many underdoped cuprates. When the damping becomes insensitive to the
$d$SC correlation length, the pseudogap region shrinks as in the
overdoped cuprates.}

\kword{antiferromagnetism, $d_{x^2-y^2}$-wave superconductivity,
self-consistent renormalization, van-Hove singularity,
high-$T_c$ superconductivity, pseudogap}

\begin{document}
\sloppy
\maketitle

\fulltext
\section{Introduction} \label{SECTION_intro}

In the underdoped region of high-$T_c$ cuprates, spin and charge
excitations widely show pseudogap phenomena in the normal metallic
states from temperatures $T_{PG}$ much higher than the superconducting
transition temperature $T_c$.  Influence of the pseudogap formation
shows up in various different experimental probes such as NMR
relaxation time, the Knight shift, neutron scattering, tunnenling,
photoemission, specific heat, optical conductivity, and DC
resistivity~\cite{RMP}.  In particular, the angle resolved
photoemission spectra
(ARPES)~\cite{LoeserShenDessauMarshallParkFournierKapitulnik1996,photoemission}
have revealed that the pseudogap starts growing first in the
single-particle excitations around $(\pi,0)$ and $(0,\pi)$ points at
$T=T_{PG}$. With decrease in  temperature, the formation of the
pseudogap gradually extends in the direction of diagonal in the
Brillouin zone such as $(\pi/2,\pi/2)$ and seems to continuously merge
into the $d_{x^2-y^2}$ gap below $T_c$.
From this observation, it is clear that the $(\pi,0)$ and $(0,\pi)$
regions in the single-particle excitation have a particular importance
in the mechanism of the pseudogap formation because the pseudogap
starts from these regions.  In fact $(\pi,0)$ and $(0,\pi)$ regions
are known from the ARPES data as the region where the quasiparticle
dispersion is very flat with strong damping in the underdoped
cuprates.  This seems to suggest that the pseudogap instability is
caused because of such flat and damped dispersion.  We call such
$(\pi,0)$ and $(0,\pi)$ momenta ``flat spot" and the region around
them ``flat shoal region" and pay a particular attention in this paper
to understand the mechanism of the pseudogap formation.  
We note that the flat dispersion around the flat spots is also
reproduced in numerical calculations~\cite{Imada98,Assaad98} and plays 
a particularly important role in the understanding of the
metal-insulator transition~\cite{Imada98,RMPsecIIF11}

A remarkable point of the pseudogap is that it appears in both of spin
and charge excitations and its gap structure in the momentum space is
the same as the superconducting $d_{x^2-y^2}$ symmetry with continuous
evolution through $T_c$.  This implies that the pseudogap phenomena
have close connection to the superconducting fluctuations.   Another
somewhat puzzling point to be stressed is that although the pseudogap
structure appears in $1/T_1T$~\cite{Yasuoka,Y124NMR,Bi2212NMR,Hggap,Hg1223NMR},
in many cases, $1/T_{2G}$ which measures ${\rm Re}
\chi(Q=(\pi,\pi),\omega=0)$ continuously increases with the decrease
in temperature with no indication of the pseudogap.   This implies
that the antiferromagnetic excitations are suppressed below $T_{PG}$
but not in a simple fashion.  A related observation is the so called
resonance peak in the neutron scattering experiments~\cite{Fong}.  A
resonance peak above the pseudogap frequency sharply grows  below
$T_c$ or $T_{PG}$.  This peak frequency $\omega^\ast$ decreases with decreasing
doping concentration, which suggests that the resonance peak may well
continuously evolve into the Bragg peak in the AFM phase.  This
strongly supports the idea that the AFM fluctuations are suppressed
around $\omega=0$ but well retained or even enhanced at a nonzero
frequency below $T_{PG}$.  These observations lead to the necessity to
consider the superconducting and antiferromagnetic fluctuations on an
equal footing and  examine their interplay in detail.

From theoretical point of view, the origin of the  pseudogap is still
controvertial~\cite{RMP2}.  It is important to understand why the
pseudogap region is so wide in the underdoped region of the high-$T_c$
cuprates.  In the weak-coupling superconductors, pairing fluctuations
are usually observed only very close to $T_c$ and the BCS mean-field
description is rather adequate. However, in the high-$T_c$ cuprates,
the coherence length is only several lattice spacings even within a
CuO$_2$ plane, and much smaller than the BCS superconductors.  Such
strong coupling nature of the pairing makes the fluctuation region
wider. In addition, the quantum degeneracy temperature (coherence
temperature $T_{coh}$ ) approaches zero if the Mott insulating state
is approached in a continuous fashion.  This necessarily leads to the
vanishing $T_c$.  In the underdoped region, such suppression of
coherence becomes stronger and it could become smaller than the energy
scale of the pairing force($\sim T_{PG}$).  With decreasing doping
concentration, if the pairing force remains finite, the separation of
$T_c$ and $T_{PG}$ is a natural consequence.  

Sometimes such suppressed $T_c$ far below $T_{PG}$ is modelled by the
Bose
condensation~\cite{Leggett80,NozieresSchmitt-Rink85,EmeryKivelson95,pseudo-gap,Hetrog9808051}.
However, this strong coupling aspect cannot be well described by a
simple Bose condensation, because the superconducting phase appears
near the strongly correlated insulator, the Mott insulator.  Because
of its correlated nature which we try to clarify in this paper, the pairing fluctuations do not
appear as simple bosons but as a complicated interplay with underlying
metal-insulator transition, antiferromagnetic fluctuations, $d$-wave
symmetry of pairing, and strong momentum dependence of single-particle
excitations persisted in the pseudogap region.  An important point
often ignored in the literature is the vertex correction, which we
discuss its effects in this paper.  

In this paper, we make effort to understand the origin of the
pseudo-gap behavior in AFM spin excitations.  We consider the problem
under the condition that both of the $d$SC and AFM fluctuations play
roles in low energy scales.  The pseudogap seen in the spin
excitations is consistently understood from the $d$SC fluctuations
and precursor effects to the $d$-wave superconductors.

One-loop renormalization group treatments of electrons
around the van-Hove singular points at the Fermi
level~\cite{Schulz87,Dzyaloshinskii96,FurukawaRiceSalmhofer9806159}
show the appearance of attractive vertices for the $d$SC channel as
well as for the AFM, though it is difficult to discuss properties
at their strong coupling fixed point from them.
Learning from this insight, in \S~\ref{SECTION_Seff}, we derive the low-energy
effective action for the relevant AFM and $d$SC fields
from the electrons around the flat spots.
In \S~\ref{SECTION_SCR}, mode-mode coupling terms are considered in
the effective action. We apply and extend the SCR (self-consistent
renormalization)
theory developed by Moriya and his coworkers to improve the RPA~\cite{SCR}.
In \S~\ref{SECTION_results}, we give numerical results and their
physical interpretations, and discuss the relation to the pseudogap
in high-$T_c$ cuprates.
We find two characteristic temperatures, $T_{PG}$ and $T_\ast$ when
the $d$SC order appears at $T=0$.  The spin pseudogap with suppression
in $1/T_1T$ appears below $T_{PG}$.  Below this temperature, the $d$SC
correlations become dominant over the AFM correlations and the
resultant decrease in the paramagnon damping takes place.  The AFM
correlations are still increasing with decrease in temperature in this 
region. Below $T_\ast$ the antiferromagnetic correlation starts
decreasing while the pairing correlations
grow as in renormalized classical regime and the superconducting
transition occurs shortly.  If the paramagnon damping is determined
from the scattering from the flat shoal region, as expected in the
underdoped region, $T_{PG}$ is substantially higher
than $T_{\ast}$.  In the region $T_{\ast}<T<T_{PG}$, the precursor to
the superconducting state develops the pseudogap structure.  When the
paramagnon damping is dominantly controlled from the region other than
the flat shoal region, $T_{PG}$ merges into $T_\ast$.
These reproduce the basic features of the experimental results.  
\S~\ref{SECTION_summary} is the summary of this paper
and the problems left for the future are also dicussed.

\section{Low-Energy Effective Action} \label{SECTION_Seff}

Under the condition that AFM fluctuations are the only important
ones in the long-length scale of the system, several
theoretical and approximate descriptions were developed.  For example,
in the Mott insulating systems, Chakravarty, Halperin and Nelson
developed a renormalization group scheme for the low-energy AFM
exciations by employing the $d\!+\!1$-dimensional nonlinear-$\sigma$
model.~\cite{ChakravartyHalperinNelson89,ChubukovSachdevYe94}.
Another quite different treatment of the AFM fluctuations in the
literature is the Gaussian approximation of the paramagnetic metallic
phase by starting from the Ginzburg-Landau-Wilson action in the
overdamped regime due to the coupling to the Stoner
excitations.~\cite{HertzKlenin74,Hertz76,Millis93} 

There exist, however, more complicated cases where other type of
fluctuations also come into play in the low energy scale.  In this
case, interplay and competition between different fluctuations
seriously affect physical properties in general with possibilities of
new critical phenomena.  The high-$T_c$ cuprates
provide such examples, where pairing fluctuations are expected to
compete with AFM fluctuations, because AFM and dSC long-ranged ordered
phases are observed close each other in the two-dimensional $T-\delta$
(temperature-hole concentration) phase diagram. 

Even when one starts from the path integral with the auxiliary fields
describing only the AFM fluctuations, in principle, all the other
fluctuations are of course also correctly taken into account if the
functional integral is performed honestly up to infinite order. In
princple, it can be done by taking account all the fluctuations around
the AFM saddle point.  It is, however, practically not easy to handle.
Physically meaningful results are rather obtained by taking account
not only of the AFM saddle point but also additional contributions
from other relevant saddle points separately. 
This procedure is justified if the overlap between fluctuations around
different saddle point is properly considered to keep away from double
counting of the degrees of freedom. 
To discuss interplay between dSC and AFM fluctuations, here, 
we treat the AFM and dSC saddle points and fluctuations around them on
equal footing. 

We consider a 2D strongly correlated electron system and assume that
AFM and pairing fluctuations are both strong as in the high $T_c$
cuprates.  The AFM fluctuations are easily taken into account in the
standard
functional-integral technique by  the Stratonovich-Hubbard
transformation for the local Coulomb repulsion, in which integrals
over auxiliary fields for the AFM order parameter appear.
The Stratonovich-Hubbard field for the AFM fluctuations is given by
SU(2) symmetric variable $\sphi$.  At the moment the microscopic
process which drives the $d$-wave pairing flucutations is not
competely understood.
Although we discuss below possible microscopic origins of the pairing
saddle point, we first introduce $U(1)$ symmetric auxiliary field
$\dphi$ on phenomenologcal grounds.  Then
the partition function represented by the functional integral over
both of the AFM and dSC auxiliary fields, $\sphi$ and $\dphi$,
respectively reads
\begin{eqnarray}
Z&=&\smint{\cal D}\adagger{\cal D}a{\cal D}\sphi
{\cal D}\bdphi{\cal D}\dphi e^{-\tilde{S}} \\
\tilde{S}&=&-\beta\sum_m\smintk(i\Omega_m-\varepsilon_{\mib k})
\adagger_\alpha(i\Omega_m,{\mib k})a_\alpha(i\Omega_m,{\mib k})
\nonumber \\
&&+\beta\sum_n\smintq\Big[
2\sqrt{|\sGamma|T}{\mib S}(i\omega_n,{\mib q})\!\cdot\!
\sphi(-i\omega_n,-{\mib q})
+i\sqrt{|\dGamma|T}\left\{\dddagger(i\omega_n,{\mib q})\dphi(i\omega_n,{\mib q})
+d(i\omega_n,{\mib q})\bdphi(i\omega_n,{\mib q})\right\}
\nonumber \\
&&+\sphi(i\omega_n,{\mib q})\!\cdot\!\sphi(-i\omega_n,-{\mib q})
+\bdphi(i\omega_n,{\mib q})\dphi(i\omega_n,{\mib q})\Big],\label{Sfb}
\end{eqnarray}
where $\adagger$ and $a$ are Grassman fields for electrons,
$\sphi$ is the three-component vector field corresponding to the spin,
and $\bdphi$ and $\dphi$ are the pairing fields creating and annihilating
a pair of electrons, respectively.
$\beta$ is the inverse temperature and $\omega_n$ and $\Omega_m$ are
bosonic and fermionic Matsubara frequencies.
The bare dispersion is described by $\varepsilon_{\bf k}$.  Here, we
take the nearest-neighbor transfer $t$ and the next-nearest-neighbor
transfer $t'$, which leads to $\varepsilon_{\bf k} = -2t({\rm cos}k_x
+{\rm cos}k_y)-4t'({\rm cos}k_x{\rm cos}k_y+1)-\mu$, where $\mu$ is
the chemical potential measured from the flat spots.
Further we have introduced the following operators corresponding to
spin, and annihilating and creating a $d_{x^2-y^2}$-wave electron
pair,
\begin{eqnarray}
{\mib S}(i\omega_n,{\mib q})&=&\sum_m\smintk
\adagger_\mu(i\Omega_m-i\omega_n,{\mib k}-{\mib q}){\mib \sigma}_{\mu\nu}
a_\nu(i\Omega_m,{\mib k}) \\
d(i\omega_n,{\mib q})&=&\sum_m\smintk
a_\mu(i\Omega_m,{\mib k})\sigma^y_{\mu\nu}
a_\nu(-i\Omega_m+i\omega_n,-{\mib k}+{\mib q})g({\mib k}) \\
\dddagger(i\omega_n,{\mib q})&=&\sum_m\smintk
\adagger_\mu(i\Omega_m-i\omega_n,{\mib k}-{\mib q})
\sigma^y_{\mu\nu}\adagger_\nu(-i\Omega_m,-{\mib k}),
\end{eqnarray}
where $\sigma_{\mu\nu}^i$ is the $i$-th component of the Pauli matrix, 
and $g({\mib k})=\cos k_x -\cos k_y$.
The term containing $\sGamma$ is obtained from the Coulomb
interaction term by the Stratonovich-Hubbard transformation for the
AFM auxiliary fields $\sphi$ while the term containing $\dGamma$
represents coupling to the pairing auxiliary field $\dphi$. 

To understand the possible origin of the term containing the dSC
fluctuations $\dphi$ and $\bdphi$, the one-loop level analyses
provide a useful insight.  On this level, in 2D systems, we first have
to consider the contribution from the quasi-particles around $(\pi,0)$
momentum, because the $(\pi,0)$ point contribution generates the most
divergent logarithmic terms due to the Umklapp scattering and also due
to the van-Hove singularity. We consider two-dimensional interacting
electron systems with the Fermi level being located near the flat
spots, namely $(\pi,0)$ and $(0,\pi)$ points. As we mentioned in \S 1,
these flat spots are not only the original van-Hove singular point but
also the points around which flat dispersions are generically observed
experimentally~\cite{LoeserShenDessauMarshallParkFournierKapitulnik1996,Gofron}
as well as numerically even when the bare band structure does not
suggest the van-Hove singularity at these
spots~\cite{Imada98,Assaad98,RMP2}.
One-loop renormalization group theories show that at low energies,
both of the $d$SC and AFM susceptibilities become divergent.  In fact
the $d$SC channel is most strongly divergent when the nesting conditon
is not
satisfied.~\cite{Schulz87,Dzyaloshinskii96,FurukawaRiceSalmhofer9806159}
It implies that the irreducible vertices for the $d$SC and AFM
channels are negative at energies low enough compared with the
ultraviolet cutoff.  Although the one-loop calculations are justified
only in the weak coupling region and it only suggests that the
fluctuations are scaled to strong coupling, it is at least clear that
we have to consider the two fluctuations explicitly in the competing
region.

After integrating out the fermions in (\ref{Sfb}), the following effective action is obtained,
\begin{eqnarray}
S &=& S^{(0)} + S_\sigma^{(2)} + S_d^{(2)}
+ S_{\sigma\sigma}^{(4)} + S_{dd}^{(4)} + S_{\sigma d}^{(4)}
\label{S} \\
S_\sigma^{(2)} &=& \beta\sum_n\smintq\schi^{-1}(i\omega_n,{\mib q})
\sphi(i\omega_n,{\mib q})\!\cdot\!\sphi(-i\omega_n,-{\mib q})
\label{S_sigma2} \\
S_d^{(2)} &=& \beta\sum_n\smintq\dchi^{-1}(i\omega_n,{\mib q})
\bdphi(i\omega_n,{\mib q})\dphi(i\omega_n,{\mib q})
\label{S_d2} \\
S_{\sigma\sigma}^{(4)} &=& \beta\uss\hspace{-10pt}\sum_{n_1,n_2,n_3}\!
\smintq_1\smintq_2\smintq_3
\sphi(i\omega_{n_1},{\mib q}_1)\!\cdot\!\sphi(i\omega_{n_2},{\mib q}_2)
\sphi(i\omega_{n_3},{\mib q}_3)\!\cdot\!\sphi(i\omega_{n_4},{\mib q}_4)
\label{S_sigma4} \\
S_{dd}^{(4)} &=& \beta\udd\hspace{-10pt}\sum_{n_1,n_2,n_3}\!
\smintq_1\smintq_2\smintq_3
\bdphi(i\omega_{n_1},{\mib q}_1)\dphi(i\omega_{n_2},{\mib q}_2)
\bdphi(i\omega_{n_3},{\mib q}_3)\dphi(i\omega_{n_4},{\mib q}_4)
\label{S_d4} \\
S_{\sigma d}^{(4)} &=& 2\beta\usd\sum_{n_1,n_2,n_3}\!
\smintq_1\smintq_2\smintq_3
\sphi(i\omega_{n_1},{\mib q}_1)\!\cdot\!\sphi(i\omega_{n_2},{\mib q}_2)
\bdphi(-i\omega_{n_3},-{\mib q}_3)\dphi(i\omega_{n_4},{\mib q}_4),
\label{S_sigmad4}
\end{eqnarray}
where
\begin{eqnarray}
\chi_\sigma (i\omega_n,{\mib q}) &=&\left(\zxi_\sigma\!{}^{-2}+({\mib q}-{\mib Q})^2+\gamma_\sigma |\omega_n|/c_\sigma^2+(\omega_n/c_\sigma)^2\right)^{-1}, \label{chi_sigma} \\
\chi_d (i\omega_n,{\mib q}) &=&\left(\zxi_d\!{}^{-2}+{\mib q}^2+\gamma_d |\omega_n|/c_d^2
+(\omega_n/c_d)^2\right)^{-1}, \label{chi_d}
\end{eqnarray}
are AFM and $d$SC dynamical susceptibilities,
${\mib q}_4=-{\mib q}_1-{\mib q}_2-{\mib q}_3$, $n_4=-n_1-n_2-n_3$,
and ${\mib Q}$ is the AFM ordering wave vector.
In this paper we restrict ourselves to the AFM ordering vector
at the commensurate value $(\pi,\pi)$,
and neglect possible long-range features of Coulomb interaction
which may lead to gapful $d$SC excitations instead of the Goldstone
mode even in the $d$SC ordered state, as in the $s$-wave SC state.
$\zxi_\sigma$ and $\zxi_d$ are the spin correlation length
and the coherence length at the mean-field level, respectively, and
the phase excitations (Higgs bosons) are not treated separately from
the amplitude modes.
The velocity of spin and pairing collective modes are
denoted by $c_\sigma$ and $c_d$. The damping constants are given by 
$\gamma_d$ and $\gamma_{\sigma}$.

We first discuss how $\zxi_\sigma$ and $\zxi_d$ in
(\ref{chi_sigma})-(\ref{chi_d}) are derived. If we employ the RPA and
$T$-matrix approximation, {\it i.e.}, the Gaussian approximation for all the
above fields, $\zxi_\sigma$ and $\zxi_d$ depend on the temperature $T$
and the chemical potential $\mu$ measured from the flat spots in the
following way:
\begin{eqnarray}
\zxi_d\!{}^{-2} &\approx&
\!1\!-\!\frac{|\dGamma|}{\sqrt{t^2-4t^\prime{}^2}}
\log\frac{E_c}{T}\log\frac{E_c}{max\{\mu,T\}} \label{RPA} \\
\zxi_\sigma\!{}^{-2} &\approx&
\!1\!-\!\frac{|\sGamma|}{t}\!\log\!\frac{E_c}{max\{\!\mu,t^\prime\!,T\!\}}\!
\log\!\frac{E_c}{max\{\!\mu,T\!\}}. \label{TMA}
\end{eqnarray}
Here $E_c$ is a ultraviolet cutoff in the energy scale of the
bandwidth and we have included coefficients of the double logarithms
into $\sGamma$ and $\dGamma$.
If electrons near the $(\pi,0)$ and $(0,\pi)$ points play dominant
role for the both ordering fluctuations, in the mean-field level,
finite-temperature transitions either to the $d$SC at $T=\zT_c$, or
the AFM at $T=\zT_N$ are expected as described by
\begin{eqnarray}
\zT_N&\approx&\left\{\begin{array}{@{\,}ll}
E_c \exp\left(-\sqrt{t/|\Gamma_\sigma|}\right)
 \\ \hspace{90pt}
\mbox{for\,\,} \zT_N>|\mu|,|t^\prime| \\
E_c \exp\left(-\frac{t}{|\Gamma_\sigma|\log E_c/|t^\prime |}\right)
 \\ \hspace{90pt}
\mbox{for\,\,} |t^\prime|>\zT_N>|\mu|
\end{array}\right. \label{T_NRPA} \\
\zT_c&\approx&\left\{\begin{array}{@{\,}ll}
E_c\exp\left(-\sqrt{\frac{\sqrt{t^2-t^\prime{}^2}}{|\Gamma_d|}}\right)
 \\ \hspace{90pt}
\mbox{for\,\,} \zT_c > |\mu | \\
E_c \exp\left(-\frac{\sqrt{t^2-t^\prime{}^2}}
{|\Gamma_d|\log E_c/|\mu |}\right)
 \\ \hspace{90pt}
\mbox{for\,\,} \zT_c < |\mu |
\end{array}\right. \label{T_cTMA}
\end{eqnarray}

We next discuss how $\gamma_\sigma$ and $\gamma_d$ are determined.
In principle, they may be determined microscopically from the
scattering of the collective modes by quasiparticles.
In this paper, however, we consider them from phenomenological
arguments. The origin of $\gamma$ is mainly from
contimnuum of the Stoner excitations and the amplitude strongly depends
on the density of states of low-energy excitations.  This low-energy
part of damping is usually absent if some kind of long-ranged order
appears, and it is also suppressed if the correlation length gets
longer. When only one type of fluctuations with the correlation length
$\xi$ exists, a plausible dependence for long $\xi$ would be
$\gamma=\zgamma \xi^{-\varphi}$.

For the choice of $\gamma$,  here, two possible cases are of interest.
In one case, AFM and $d$SC fluctuations are coupled with
quasiparticles excited far from the flat spots $(\pi,0)$ and
$(0,\pi)$, and they dominate dampings of the above collective
fluctuations.  In this case we expect $\varphi=0$.
In the other case, the main source of the damping is ascribed to the
quasiparticles around the flat spots.  Then we expect $\varphi>0$.

In \S 4, we give the calculated results of the pairing correlation
length, the spin correlation length and the NMR spin-lattice
relaxation rate for $\varphi=0$ and $1$. When only the AFM spin
fluctuations exist, a corssover between the dynamical exponent $z=2$
at low temperatures and the $z=1$ at high temperatures in the
disordered side near the critical point has been treated by assuming
$\varphi=0$ by Sachdev {\it et al.}~\cite{SachdevChubukovSokol95}.
Recently, Schmalian has discussed the case, $\varphi=1$ by
the renormalization group method and has obtained a crossover between
the behavior with $z=1$ at the disordered side near the critical point and
the $z=2$ behavior far from the critical region~\cite{Schmalian9810041}.
In this paper, we show that in the presence of another strong
instability, the system shows properties significantly different from
the case with only one type of fluctuations. We discuss this problem
further in \S 4.

Since other parameter values $c_{\sigma}$ and $c_d$ do not have such
singular dependences, in this paper, we neglect their dependence on
$t^\prime$, $\mu$ and $T$. For example, we can safely regard the
velocities, $c_\sigma$ and $c_d$ as constants, since the velocities
remain finite at the transition point.

Next we proceed to the perturbative evaluation of quartic terms.
The detailed calculations of the coupling constants are presented in
Appendix~\ref{APPENDIXcouplings}.
Roughly speaking, the AFM-$d$SC coupling constant turns out to be
positive when $0\le\mu < t-2t^\prime$, $0\le -\mu<t+2t^\prime$, or
$T>|\mu|$ are satisfied. The other couplings are always positive.
$u_{dd}$ is the most divergent at the lowest energy.
It diverges as $\ln E_c/Max\{T,\mu\}$ when $T\to 0$.
This leads to nothing but the breakdown of the expansion of $S_{eff}$
in terms of the bare Green function for small enough $|\mu|$.
We, then, introduce phenomenological constant values for the mode-mode
couplings as mentioned in \S 4.
In short, our theory displayed below corresponds to a treatment of a
phenomenological spin-$d$-wave pairing model.

Note that $\pi$-triplet fluctuations can be suppressed when the $SO(5)$
symmetry~\cite{Zhang97science,Zhang98preprint} is absent.
In fact, recently Furukawa {\it et al.} have argued
within the g-ological one-loop renormalization group treatment that
for the $(\pi,0)$ contributions at the Fermi level,
this is the case even when the strengths of the divergences of
AFM and $d$SC fluctuations are the same, {\it i.e.},
when the next-nearest neighbor hopping $t'$ is equal
to zero~\cite{FurukawaRiceSalmhofer9806159}.
Therefore we assume that the suppressed $\pi$-triplet fluctuations may 
be ignored.

\section{Self-Consistent Renormalization} \label{SECTION_SCR}

\begin{figure}[t]
\begin{center}
\epsfxsize=7.8cm
$$\epsffile{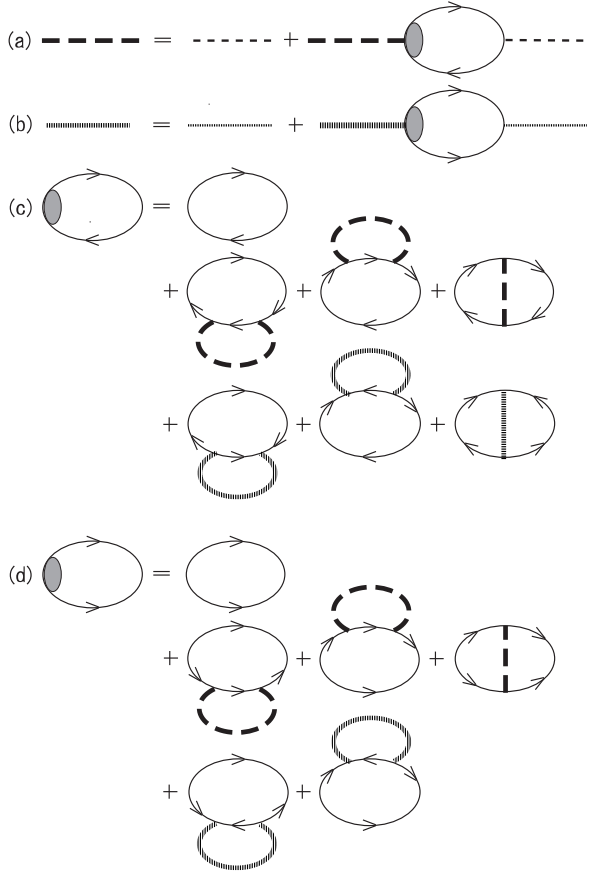}$$
\fulltext
\vspace{-0.7cm}
\caption{Diagrammatic representation of the present theory.
Solid lines denote electronic Green function while
thick (thin) long- and short-dashed lines are
renormalized (unrenormalized) AFM and $d$SC fluctuations.
The diagrams (a) and (b) represent the Dyson equations for the AFM and $d$SC fluctuations,
respectively.
The diagrams (c) and (d) are the self-energies for AFM and $d$SC fluctuations.
We have kept terms up to the one-loop order in the perturbation expansion
of them.}
\end{center}
\label{diagram}
\end{figure}

Next using the effective action obtained above, we start
renormalization process for the mode-mode coupling terms.  This is a
similar procedure to the SCR theory developed by Moriya and
coworkers~\cite{SCR}.
In our case the mode-mode coupling terms consist of those between AFM
and AFM fluctuations, $d$SC and $d$SC fluctuations, and AFM and $d$SC
fluctuations. Following the SCR scheme, we have considered the Feynmann
diagrams represented in Fig.~\ref{diagram}.
\fulltext
In the usual spin flucutation theories, diagrams containing $d$SC
fluctuations are absent. 
Here we consider the both AFM and dSC fluctuations to understand their
interplay.
There exists some previous attempts related to this line of approach.
For example, the coupling of spins to pairing fluctuations was
discussed in the calculation of the uniform spin susceptibility near
the superconducting transition~\cite{RanderiaVarlamov94}.
However this approach did not consider AFM fluctuations.  It also did
not take an explicit account of electrons around the $(\pi,0)$ point. 

The calculated results in fact depend on the choice of the parameter
values in the effective action Eq.~(\ref{S}).
Among other parameter values in (2.6)-(2.13), only those representing
the bare distance from the critical point, $\zxi_\sigma{}^{-2}$ and
$\zxi_d{}^{-2}$ behave singularly at $T=0$ even for finite values of $\mu$,
and they are still most divergent with logarithmic accuracy.
Then they crucially determine the qualitative aspects through the
dependence on the choices of $t^\prime$, $\mu$ and $T$. 

Consequently, the role of the self-consistent renormalization
procedure is reduced to the determination of $\xi_\sigma$ and $\xi_d$ from
\begin{eqnarray}
\xi_\sigma^{-2}&=&\zxi_\sigma{}^{-2}+
\!\int_0^{2E_c}\!\frac{d\omega}{\pi}\smint\!\frac{d^2{\mib k}}{(2\pi)^2}
\coth\frac{\omega}{2T}
\left[u_{\sigma\sigma}{\rm Im}\chi_\sigma(\omega,{\mib k})
+u_{\sigma d}{\rm Im}\chi_d(\omega,{\mib k})\right]
\label{SCRs} \\
\xi_d^{-2}&=&\zxi_d{}^{-2}+
\!\int_0^{2E_c}\!\frac{d\omega}{\pi}\smint\!\frac{d^2{\mib k}}{(2\pi)^2}
\coth\frac{\omega}{2T}
\left[u_{\sigma d}{\rm Im}\chi_\sigma(\omega,{\mib k})
+u_{dd}{\rm Im}\chi_d(\omega,{\mib k})\right].
\label{SCRd}
\end{eqnarray}

Our main results obtained numerically will be discussed in the
following section.
Here, we comment on the characteristics of the present SCR theory for
the mode-mode couplings.
First, as in the SCR theory of spin fluctuations in 2D,
the system can be ordered only at $T=0$.
It agrees with the Mermin-Wagner theorem~\cite{MerminWagner66} for the
AFM order.
For the $d$SC, the K-T transition at nonzero temperatures is not
possible in our formalism in contrast with what should be.
This implies that the treatment of the gauge fluctuations in the SCR
theory is inadequate for the description of the K-T transition at
$T_{KT}$.
However, once all the higher energy degrees of freedom are integrated
out, the renormalized superfluid stiffness determines the temperature
scale where the pairing correlation length starts growing strongly.
Since $T_{KT}$ is of the order of the stiffness
according to the K-T theory~\cite{K-Ttransition},
$T_{KT}$ should occur close to this crossover temperature $T_\ast$
in our theory, below which the spin correlation length starts
decreasing.  In our analysis, we take this temperature
scale as the signature of the K-T transition.

Another important point is that though the Gaussian approximation for
the electrons around the $(\pi,0)$ point predicts the coexistence of
the AFM and the $d$SC, as discussed in \S~\ref{SECTION_Seff},
as well as mean-field arguments~\cite{MurakamiFukuyama98},
the inclusion of vertex corrections due to the coupling between AFM
and $d$SC modes makes their coexistence difficult. 
It is simpliy because the coupling between AFM and $d$SC modes are
repulsive for realistic cases. 
The coexistence appears only at $t^\prime\!=\!\mu\!=\!0$.
The further conditions for their stability are given by
$|\Gamma_d|<|\Gamma_\sigma|u_{\sigma d}/u_{\sigma\sigma}$
and $|\Gamma_\sigma|>|\Gamma_d|u_{\sigma d}/u_{dd}$ for the AFM,
and $|\Gamma_d|>|\Gamma_\sigma|u_{\sigma d}/u_{\sigma\sigma}$
and $|\Gamma_\sigma|<|\Gamma_d|u_{\sigma d}/u_{dd}$ for the $d$SC,
as shown in Appendix~\ref{APPENDIXnest}.
A quite different type of coexistence of dSC and AFM may in principle
be possible if different portion of the Fermi surface than $(\pi,0)$
and $(0,\pi)$ contributes to the AFM order while $d$SC order is
realized only through contributions around $(\pi,0)$ part.   Since we
do not seriously treat contributions from $(\pi/2,\pi/2)$ part, this
possibility is beyond the scope of this paper.

In what follows, we concentrate on the cases where the $d$SC ground
state is realized at $T=0$ while substantial AFM fluctuations also
persist.  In our analyses, we take $\Gamma_d\!=\!\Gamma_\sigma$.
Then the system shows the competing fluctuations of the AFM and the
$d$SC in the normal metallic (but not necessarily Fermi-liquid-like)
phase below the mean-field transition temperature.

\fulltext
\section{Results} \label{SECTION_results}

Before showing results, we discuss how the phenomenological parameters 
are determined.
For comparison with experiments of hole-doped high-$T_c$ cuprates
shown below, we take the following rough estimates; $t=0.25$eV from
photoemission data of
YBa${}_2$Cu${}_3$O${}_{7-x}$~\cite{MassiddaYuFreeman87-1,SiZhaLevinLu93},
$E_c=t$, $c_\sigma=0.5t$ and $\zgamma_\sigma=t$.
Hereafter we take the lattice spacing as the length scale and $t$
as the energy unit.
Above choices of $\zgamma_\sigma$ and $c_\sigma$ are taken similar to
those obtained by fitting experimental data of $1/T_1T$ and $T_{2G}$ of
YBa${}_2$Cu${}_3$O${}_{7-x}$ at high temperatures with the staggered
susceptibility of the form (\ref{chi_sigma}).~\cite{MonthouxPines94}
At the moment, some uncertainty exists in determining the parameter
values describing $d$SC fluctuations.
For simplicity and definiteness, we take $c_d=c_d$ and
$\zgamma_d=\zgamma_\sigma$.
We assume that $t'$ and $\mu$ are small compared with those obtained
by tight-binding band fittings~\cite{SiZhaLevinLu93} with the photoemission
results~\cite{YBCO1236.9ARPES}, and take comparable values to $T_c$,
so that contributions from the flat shoal regions are dominant.
We note that experimentally observed Fermi surface may be reproduced
by such parameters due to the strong correlation
effects~\cite{TsunetsuguImadaU}.
For $\zxi_{\sigma,d}$, we adopt the one-loop results (\ref{RPA}) and
(\ref{TMA}) with proportionality constants determined so that the spin
and pairing correlation lengths at $T=1$ are a unity.
Further we take the values of fermionic vertices $\sGamma=\dGamma$
which predict $\zT_N$ of order of the temperature below which the
system enters a critical region.
Unfortunately, it is rather difficult to reliably estimate the
renormalized values of the mode-mode couplings in the available framework.
In this paper, we leave $\sGamma=\dGamma$ and the mode-mode couplings
as adjustable parameters and pose a question whether the experimental
results can be totally reproduced within reasonable choices of their
values. Particularly, the following conditions have to be satisfied at 
least to make more detailed comparisons with the experimental data:
First, the onset temperature $T_\ast$ where correlation
length $\xi_d$ starts increasing rapidly should be close to $T_c$ in
real materials. Second, $\xi_\sigma$ at $T=T_\ast$ is similar to that 
obtained in the experiments in each case. Third, the crossover
temperature $T_{PG}$ obtained in the case $\varphi=1$ at which $1/T_1T$
starts decreasing gives a pseudogap temperature obtained experimentally.
%

\begin{figure}[htb]
\begin{center}
\epsfxsize=7.8cm
$$\epsffile{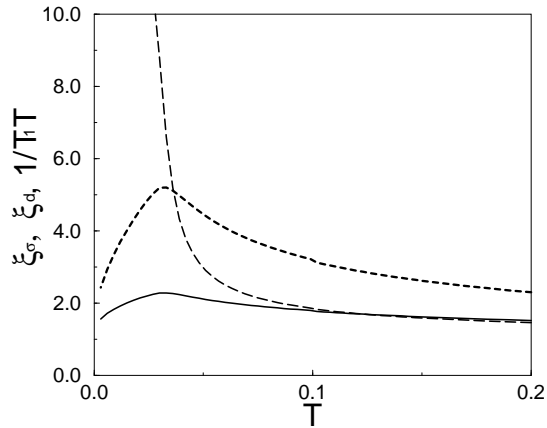}$$
\fulltext
\vspace{-0.7cm}
\caption{The calculated spin correlation lengths (solid lines),
$1/T_1T$ (dashed lines) normalized by its value at $T=1$,
$d$SC correlation lengths (long-dashed lines),
We have taken $|\sGamma|=|\dGamma|=0.15$, $\uss=2.6$, $\usd=1.2$,
$\udd=1.8$, $\mu=0.03$ and $t'=-0.03$,
which correspond to YBa${}_2$Cu${}_3$O${}_7$.}
\label{Figxi-phi0}
\end{center}
\end{figure}
\fulltext
First, for the case $\varphi=0$, we show the case where the parameter values
correspond to YBa${}_2$Cu${}_3$O${}_7$~\cite{Yasuoka}.
We plot the calculated $d$SC correlation length $\xi_d$, $1/T_1T$ and
spin correlation length $\xi_\sigma$ in Fig.~\ref{Figxi-phi0}. 
As previously mentioned, on this level of approximations, the nature
of the original Fermi surface determines whether the ground state is
AFM or $d$SC, through the bare irreducible particle-hole and
particle-particle susceptibilities.
For finite values of $t^\prime$ or $\mu$, the ground state is $d$SC in
the one-loop results.
In this case, the spin correlation length first increases down to the
temperature $T_\ast$  and then decreases with further decrease in
temperature. The crossover temperature $T_\ast$ depends on $t^\prime$,
mode-mode couplings and other parameters in the effective action.
It makes a crossover between the regime $T<T_\ast$ dominated
by the $d$SC renormalized classical fluctuations and the thermally
fluctuating regime $T>T_\ast$.

The spin-lattice relaxation rate $T_1$ of a ${}^{63}$Cu nuclei,
after a proper subtraction of the contribution from the uniform part,
is evaluated approximately as
\begin{eqnarray}
1/T_1T&\propto&\lim_{\omega\to 0}\smint\!d{\mib k}{\rm Im}\chi_\sigma(\omega,{\mib k})/\omega \nonumber \\
&\propto&\xi_\sigma^2\gamma_\sigma/c_\sigma^2. \label{1/T_1T}
\end{eqnarray}
%
Here we have not considered the ${\bf k}$-dependence of the nuclear
form factor seriously, because it does not alter the basic feature.
From (\ref{1/T_1T}), we see that $1/T_1T$ has basically the same
temperature dependence as $\xi_\sigma^2$.
By taking $T_c$ comparable to $T_\ast$, our present result for
$\varphi=0$ suggests that overdoped cuprates which show
$T_{PG}\sim T_\ast\sim T_c$ as in Tl${}_2$Ba${}_2$CuO${}_6$,
or HgBa${}_2$CuO${}_{4+\delta}$~\cite{Hg1201NMR} are also classified
as this class, besides optimally doped systems such as
YBa${}_2$Cu${}_3$O${}_7$~\cite{Yasuoka}.
We conclude that our calculated results with $\varphi=0$ are totally
consistent with the experimental results of optimally as well as
overdoped cuprates.
Even in the underdoped region,
La${}_{2-x}$Sr${}_x$CuO${}_4$~\cite{LSCONMR} seems to belong to this category.
Recent ARPES data of La${}_{2-x}$Sr${}_x$CuO${}_4$ clarified
exceptionally strong damping in the region $(\pm\pi/2,\pm\pi/2)$.~\cite{LSCOARPES}
This may well be reproduced by $\varphi=0$ class. The absence of clear 
pseudo gap in the magnetic response in this compound is consistent
with this interpretation.

Next we consider $\varphi>0$ cases. Contrary to the case $\varphi=0$,
one has to be careful in determining the scaling form of the damping rates.
If only AFM spin fluctuations exist, it can follow the simple scaling form
$\gamma_\sigma\propto\xi_\sigma^{-\varphi}$.  When we take
$\varphi=1$, the $z=1$ scaling holds within the tree level.
However, when we treat complicated fluctuations containing both AFM and $d$SC
characters tightly coupled each other through $S_{\sigma d}^{(4)}$,
the correct theory should involve two different scales $\xi_\sigma$ and $\xi_d$.
Here, note that the lower energy scale (or longer length scale)
always determines the damping rates of the modes: physically,
the strongest fluctuations affect the fermionic low-energy spectral weights,
which dominate the mode dampings.
To take into account this feedback effect on the mode dampings,
we introduce the following phenomenological relation:
\begin{equation}
\gamma_{\sigma,d}=2\zgamma_{\sigma,d}/(\xi_\sigma^{\varphi}+\xi_d^{\varphi}).
\label{damp_z=1whole}
\end{equation}
In terms of bosonic excitations,
the relaxation times of collective modes should be determined
by the time necessary to propagate the scale of the longest
correlation length, because the damping is not effective as far as the 
excitations are propagating inside such an ordered domain.
Thus we take $\varphi=1$ here.

\begin{figure}[htb]
\begin{center}
\epsfxsize=7.8cm
$$\epsffile{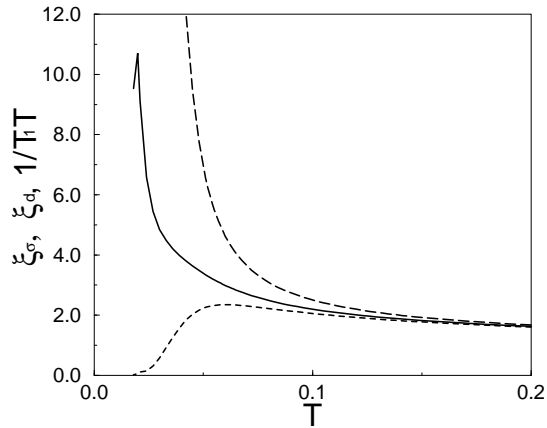}$$
\fulltext
\vspace{-0.7cm}
\caption{The results for the spin correlation lengths (solid lines),
$1/T_1T$ (dashed lines), and the $d$SC correlation lengths
(long-dashed lines) in the $\varphi=1$ case.
We have taken $|\sGamma|=|\dGamma|=0.7$, $\uss=3.14$, $\udd=3.05$,
$\usd=1.0$, $\mu=0.02$ and $t'=-0.02$, which correspond to
YBa${}_2$Cu${}_3$O${}_{6.63}$.
As in Fig.~\ref{Figxi-phi0}, the data of $1/T_1T$ are normalized by
its value at $T=1$.}
\label{Figxi-phi1}
\end{center}
\end{figure}
\begin{figure}[bht]
\begin{center}
\epsfxsize=7.8cm
$$\epsffile{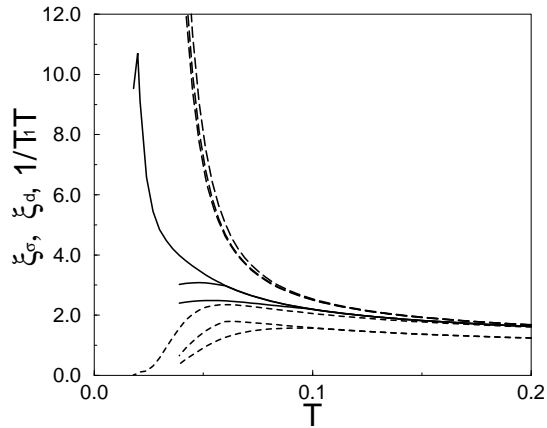}$$
\fulltext
\vspace{-0.7cm}
\caption{Spin correlation lengths (solid lines) and $1/T_1T$ (dashed
  lines) calculated in the case of $\varphi=1$ for $|t'|=0.02$, $0.06$
  and $0.1$ from top to bottom curves, and $d$SC correlation lengths
  (long-dashed lines) from bottom to top curves.
  We have taken $|\sGamma|=|\dGamma|=0.7$, $\uss=3.14$, $\udd=3.05$,
  $\usd=1.0$, $\mu=0.02$. It suggests that as $|t'|$ become larger,
  $T_\ast$ and $T_{PG}$ increase.}
\label{Figxi-phi1_tprime}
\end{center}
\end{figure}
\begin{figure}[htb]
\begin{center}
\epsfxsize=7.8cm
$$\epsffile{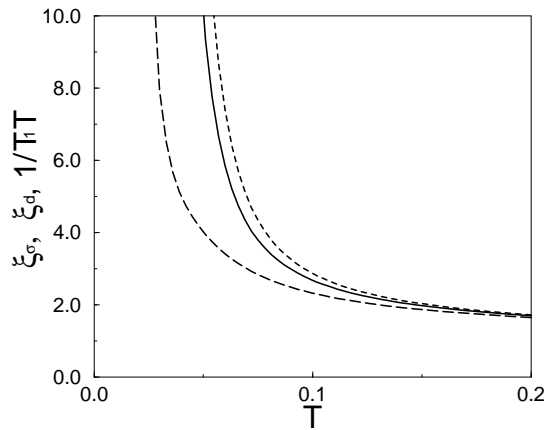}$$
\fulltext
\vspace{-0.7cm}
\caption{The results for the spin correlation lengths (solid lines),
$1/T_1T$ (dashed lines), and the $d$SC correlation lengths
(long-dashed lines) in the $\varphi=1$ case. We have taken
$|\sGamma|=|\dGamma|=0.7$, $\uss=3.0$, $\udd=3.1$, $\usd=1.0$, $\mu=0$
and $t'=0$, so that the spin correlation length shows a stronger
divergence than the $d$SC correlation length.}
\label{Figxi-phi1_co}
\end{center}
\end{figure}

\fulltext
The calculated $\xi_\sigma$, $\xi_d$ and $1/T_1T$ are shown
in Fig.~\ref{Figxi-phi1} for the parameter values corresponding
to YBa${}_2$Cu${}_3$O${}_{6.63}$.
As in $\varphi=0$ cases, with the decrease in temperature, the spin
correlation length first increases,
reaches its maximum value and then decreases due to the quantum
fluctutaions introduced by the renormalized-classical $d$SC
fluctuations.
This temperature, which we refer to as $T_\ast$, is mainly determined
from $t'$ and the mode-mode coupling constants.
On the other hand, $1/T_1T$ behaves in a different manner
from $\varphi=0$ cases.
When the temperature is lowered, $\xi_\sigma$ and $\xi_d$ both first
gradually increase.  In this region, $1/T_1T$ also gradually
increases.  With further decrease in temperature, $\xi_d$ starts
growing quicker than $\xi_\sigma$ and then $\gamma_{\sigma}$ starts
decreasing much quicker than $\xi_\sigma$ at $T_{PG}$. This
drives spin excitations from overdamped into underdamped region, and
forms a spin-excitation peak at a finite frequency $\omega=\omega^*$
in $S(Q,\omega)$.  Namely, the spectral weight start transferring from
$\omega=0$ to the peak region around $\omega^{*}$, which makes
decrease in  $1/T_1T$.  A similar crossover was previously obtained
in a numerical calculation near the quantum transition point between
$d$SC and AFM ordered phases~\cite{Assaad96,Assaad97,Assaad97b}.  At
lower temperatures characterized by $T_\ast$, the $d$SC fluctuations
go into the renormalized classical regime, which signals the decrease in
$\xi_{\sigma}$.  
We again interpret $T_{\ast}$ as the rough estimate of $T_c$.
These properties are also similar to experimental data in underdoped
cuprates with a pseudo spin gap, such as
YBa${}_2$Cu${}_4$O${}_8$~\cite{Y124NMR} and
Bi${}_2$Sr${}_2$Ca${}_1$Cu${}_2$O${}_8$~\cite{Bi2212NMR}.
In the present framework, the crossover $T_{PG}$ increases with
increasing $t'$, because it accelerates the dominance of $d$SC
correlations over AFM, which induces reduction of $\gamma_{\sigma}$.
Figure~\ref{Figxi-phi1_tprime} indeed shows that $T_{PG}$ increases with $t'$.
Figure~\ref{Figxi-phi1_co} shows the calculated results
for $\varphi=1$ and $\mu=t'=0$ with $u_{\sigma\sigma}$ less than $u_{dd}$,
so that $\xi_\sigma$ shows a stronger divergence than $\xi_d$
and the ground state is very close to the quantum transition point
between AFM and $d$SC.

\fulltext
Irrespective of the damping rate exponent $\varphi$, low-energy spin fluctuations are in general suppressed due to the $d$SC fluctuations around its ordered state.
However, the manner of suppression is different between  $\varphi=0$ and $1$,
{\it i.e.}, between the case that low-energy fermions coupled to collective
modes exist and the case that those are absent.

The peak structure in $S(Q,\omega)$ around $\omega^*$ reproduces some
qualitative feature in the resonance peak observed
experimentally~\cite{Fong,Y1236.69neutron}.  In our treatment, $\omega^*$
is selfconsistently determined from the competition between $d$SC and
AFM and the value $\omega^*$ is characterized by  the $d$SC gap
amplitude.  The AFM fluctuations are pushed out from the region lower
than $\omega^*$ due to the dSC pseudogap formation. 

Experimentally, the resonance peak appears to be sharpened with the
increasing $d$SC correlation length $\xi_d$ and the intensity of the
peak remarkably increases  below the superconducting $T_c$.
This implies a rapid growth of the AFM correlation length $\xi_\sigma$.
On the other hand, $\omega^*$ does not strongly depend on the
temperature. The frequency $\omega^*$ depends only on the doping
concentration.  It in fact decreases only when the doping
concentration is decreased.  A
similar feature is also seen numerically near the AFM-$d$SC
transition~\cite{Assaad97,Assaad97b}. This feature is beyond the scope
of this paper, because, in our framework of (\ref{chi_sigma}),
$\omega^\ast$ has to be proportional to $\xi_\sigma^{-1}$ for small
$\gamma_\sigma$.  To realize rapidly increasing correlation length
with a fixed finite $\omega^\ast$, we need to modify the assumed form
(\ref{chi_sigma}) as discussed before~\cite{Assaad97}.
To reproduce the temperature dependence, it will be necessary to
correctly take into account dominat incoherent part in addition to the 
coherent response, although our susceptibility (\ref{chi_sigma})
overlooks such a subtlety.

\section{Conclusion} \label{SECTION_summary}

The pseudogap region observed in the high-$T_c$ is reproduced as the
region with enhanced $d$SC correlations.  The experimental results
for the magnetic excitations are consistently explained from
precursor effects for the superconductivity. 

From the electrons around the flat spots, $(\pi,0)$ and $(0,\pi)$, we
have considered the low-energy effective action for the auxilary
fields for the AFM and $d$SC. 
The one-particle irreducible vertices for both channels have been
taken negative values, as suggested by one-loop renormalization group
studies~\cite{FurukawaRiceSalmhofer9806159,Schulz87,Dzyaloshinskii96}.
Next, we have obtained solutions of the self-consistent one-loop
renormalization.
The inclusion of vertex correction in 2D suppresses finite-temperature
transitions.  This correctly reproduces the absence of
antiferromagnetic transition at nonzero $T$ in 2D.   The
superconducting Kosterlitz-Thouless-Berezinski transition is also
absent in this approximation in contrast to what should be.  This is
just an artifact due to neglect of the phase modes of the $d$SC order
parameter.  In this paper, we take  $T_\ast$ as a rough estimate of
$T_c$, where $T_\ast$ is defined as the onset temperature for the
renormalized classical regime of the $d$-wave pairing.  Below this
temperature, spin correlation length starts decreasing. 
Because of the repulsive coupling between $d$SC and AFM modes,
the coexistence of the $d$SC and the AFM orders does not take place
except when both degrees of freedom are nearly symmetric as explained
in Appendix~\ref{APPENDIXnest}.

The pseudogap formation clarified from the interplay of AFM and $d$SC
is summarized as follows:
When either the Fermi level $\mu$ measured from the flat spot or the
next nearest-neighbor hopping $t'$ is nonzero, the $d$SC ground state
is realized. Only in a special case of $\mu=t'=0$, $d$SC, AFM or
coexistent ground state can be realized depending on the values of the
mode-mode couplings as discussed in Appendix~\ref{APPENDIXnest}.
When the $d$SC correlation grows faster, the process of the pseudogap
formation depends on the damping exponent $\varphi$.  If $\varphi=0$,
namely the damping $\gamma$ does not depend on the $d$SC correlation
length,  the spin correlation length, $\xi_{\sigma}$ and
$1/({}^{63}T_1T)$ both reaches its maximum value at $T=T_\ast$ and
then decreases as the temperature decreases.
On the other hand, when the mode dampings are suppressed
near the transition, {\it e.g.}, $\varphi=1$,
$1/(T_1T)$ shows a faster decrease at $T_{PG}(>T_\ast)$ while $\xi_{\sigma}$ continues to increase until $T_{\ast}$.
The pseudogap temperature $T_{PG}$ increases as $t^\prime$ increases
and it tends to decrease when the nesting effect of the Fermi surface
around the flat spot becomes stronger.

In comparison with experiments in the cuprates, the results
for $\varphi=0$ reproduce the feature obtained in the optimally doped and
overdoped high-$T_c$ cuprates, such as
YBa${}_2$Cu${}_3$O${}_7$~\cite{Yasuoka} and Tl${}_2$Ba${}_2$CuO$_6$ and
HgBa${}_2$CuO${}_{4+\delta}$~\cite{Hg1201NMR},
and even some underdoped cases as La${}_{2-x}$Sr${}_x$CuO${}_4$~\cite{LSCONMR}.
For this category, we speculate that quasi-particle excitations around
$(\pi/2,\pi/2)$ point mainly contribute to the finite damping of AFM
fluctuations. 

On the other hand, in many underdoped high-$T_c$ cuprates,
$1/(T_1T)$ starts decreasing at $T_{PG}$ above $T_c$,
while the spin correlation length measured by $T_{2G}$ shows the
Curie-Weiss behavior down to $T_c$~\cite{Yasuoka,Y124NMR,Bi2212NMR,Hg1223NMR}.
This pseudogap behavior is reproduced by our $\varphi=1$ case,
if we take $T_\ast$ as the superconducting transition temperature.
In addition, the qualitative similarity between our results for
$\varphi=1$ and the experimental results suggests that the damping of
the AFM and $d$SC collective modes decreases in the pseudogap regime.
It means that low-energy fermions around the flat spots mainly
contribute to the damping.  This is consistent with the strong damping
of quasiparticle around the flat spot observed experimentally in the
underdoped region. This flat dispersion is a consequence of the
universal character of the metal-insulator transition.

We clearly need further studies for a more complete understanding of
the pseudogap in the high-$T_c$ cuprates. After the present work,
several important steps still remain.  In the next step from the
present work, a more systematic renormalization group treatment would
be interesting~\cite{RGtheory}. We
have concentrated on the single-particle excitations only around the
flat spots, $(\pi,0)$ and $(0,\pi)$ assuming their crucial importance
in the mechanism of the pseudogap formation.  However, in the one-loop
level, the origin of the flat dispersion and strong damping in this
momentum region is not fully understood.  Experimentally
the flatness and damping strength appear much more pronounced than the
expectation from the one-loop analyses. Numerical analyses also
support that this remarkable momentum dependence around the flat spots
is due to strong correlation effects.  We have to calculate
self-energy corrections as well as the vertex corrections in a
selfconsistent fashion to clarify the profoundness of such correlation
effects.  This is clearly the step beyond the one-loop level.  Near
the metal-insulator transition point, vanishing quasiparticle weight
has to be led from a selfconsistent treatment.  This will contribute to
clarify how the pairing channel appears and how the flat spots are
destabilized to the paired singlet.  We also note that the dominance
of the incoherent weight over the quasiparticle weight in the
single-particle excitations near the metal-insulator transition may
require a serious modification in the derivation of the AFM and $d$SC
susceptibilities.  The Curie-Weiss type form for the dynamic spin
susceptibility needs to be reconsidered ~\cite{RMP2}, because the spin
susceptibility is also determined mainly from the incoherent part of
the single-particle excitations which we have not considered at all in
this paper.   In addition, we have not considered effects of
quasiparticles excitated far from the flat spots.  In this context, a
more comprehensive analyses are required to see the momentum dependence.

\acknowledgement
S. O would like to thank N. Furukawa for useful discussion.
The work was supported by "Research for the Future" Program from
the Japan Society for the Promotion of Science under the grant number
JSPS-RFTF97P01103.

\appendix
\section{Calculation of the coupling constants}
\label{APPENDIXcouplings}

Here we give the quartic coupling constants at the lowest energy and the
wave vector measured from the ordering one.
If we take $g({\mib k})={\rm sgn}(\cos k_x - \cos k_y)$,
after straightforward calculations one obtains
\begin{eqnarray}
u_{\sigma\sigma}&=&\frac{\Gamma_\sigma^2}{2(2\pi)^4}T^3\sum_m\smint\! d^2{\mib k}
G(i\Omega_m,{\mib k})^2G(i\Omega_m,{\mib k}-{\mib Q})^2 \\
&=&\frac{\Gamma_\sigma^2}{2(2\pi)^4}T^2\smint\! d^2{\mib k}
\frac{1}{(\varepsilon_{\mib k}-\varepsilon_{{\mib k}-{\mib Q}})^2}
\Big[\frac{\tanh\varepsilon_{\mib k}/2T-\tanh\varepsilon_{{\mib k}-{\mib Q}}}
{\varepsilon_{\mib k}-\varepsilon_{{\mib k}-{\mib Q}}}
-\frac{1}{2T}\sech^2\varepsilon_{\mib k}/2T\Big]
\label{u_sigma-sigma} \\
u_{dd}&=&\frac{\Gamma_d^2}{2(2\pi)^4}T^3\sum_m\smint\! d^2{\mib k}
G(i\Omega_m,{\mib k})^2G(-i\Omega_m,-{\mib k})^2 \\
&=&\frac{\Gamma_d^2}{8(2\pi)^4}T^2\smint\! d^2{\mib k}\varepsilon_{\mib k}^{-2}
\Big[\varepsilon_{\mib k}^{-1}\tanh\varepsilon_{\mib k}/2T
-\frac{1}{2T}\sech^2\varepsilon_{\mib k}/2T\Big]
\label{u_d-d} \\
u_{\sigma d}&=&-\frac{\Gamma_\sigma\Gamma_d}{2(2\pi)^4}T^3\sum_m\smint\!
d^2{\mib k}G(i\Omega_m,{\mib k})G(-i\Omega_m,-{\mib k}) \nonumber \\
&&\left\{G(i\Omega_m,{\mib k}-{\mib Q})G(-i\Omega_m,-{\mib k})
-G(\Omega_m,{\mib k}-{\mib Q})G(-i\Omega_m,-{\mib k}+{\mib Q})\right\} \\
&=&\frac{\Gamma_\sigma \Gamma_d}{8(2\pi)^4}T^2\smint\! d^2{\mib k}
\Big[-\frac{\varepsilon_{\mib k}^{-1}\tanh\varepsilon_{\mib k}/2T
           -\varepsilon_{\mib k-Q}^{-1}\tanh\varepsilon_{{\mib k}-{\mib Q}}/2T}
           {\varepsilon_{\mib k}^2-\varepsilon_{{\mib k}-{\mib Q}}^2}
\nonumber \\
&&     -\frac{1}{\varepsilon_{\mib k}+\varepsilon_{{\mib k}-{\mib Q}}}
    \big(\varepsilon_{\mib k}^{-2}\tanh\varepsilon_{\mib k}/2T
  -\frac{\varepsilon_{\mib k}^{-1}}{2T}\sech^2\varepsilon_{\mib k}/2T\big)
\Big].
\label{u_sigma-d}
\end{eqnarray}
$u_{\sigma\sigma}$ and $u_{dd}$ are always positive.
$u_{\sigma d}$ is positive when $0\le -\mu < t+2t^\prime$,
$0\le \mu<t-2t^\prime$, or $T>|\mu|$ holds approximately.
Otherwise it is negative. 
Consequently, it is positive for typical Fermi surface of high-$T_c$ cuprates.
They are finite at least at finite temperatures, thus, at the energy cutoff.

\section{Phase diagram at $T=0$ for nested cases}
\label{APPENDIXnest}

In this appendix, we discuss what type of ground states is
established in the presence of positive $\usd$. We, however,
regard that $\usd$ is small enough so that $\uss\udd>\usd^2$.

To find the lowest excitation spectrum, it is convenient to rewrite
(\ref{SCRs}) and (\ref{SCRd}) in the following expressions,
\begin{eqnarray}
\xi_\sigma^{-2}&=&\zxi_\sigma{}^{-2}+\frac{\usd}{\udd}
(\xi_d^{-2}-\zxi_d{}^{-2})+(\uss-\frac{\usd^2}{\udd})\!
\int_0^{2E_c}\!\frac{d\omega}{\pi}\!\int\!\frac{d^2{\mib k}}{(2\pi)^2}
{\rm Im}\chi_\sigma(\omega,{\mib k})\coth\frac{\omega}{2T}
\label{SCRs2} \\
\xi_d^{-2}&=&\zxi_d{}^{-2}+\frac{\usd}{\uss}(\xi_\sigma^{-2}-\zxi_\sigma{}^{-2})
+(\udd-\frac{\usd^2}{\uss})\!\int_0^{2E_c}\!\frac{d\omega}{\pi}\smint\!\frac{d^2{\mib k}}{(2\pi)^2}
{\rm Im}\chi_d(\omega,{\mib k})\coth\frac{\omega}{2T}.
\label{SCRd2}
\end{eqnarray}
When $t'\ne 0$ or $\mu\ne 0$,
within weak coupling arguments (\ref{RPA}) and (\ref{TMA}),
$d$SC ordered state is realized at $T=0$.
It is because at sufficiently low temperatures,
$\zxi_\sigma{}^{-2}+(\xi_d^{-2}-\zxi_d{}^{-2})\usd/\udd$ becomes positive,
while $\zxi_d{}^{-2}+(\xi_\sigma^{-2}-\zxi_\sigma{}^{-2})\usd/\uss$
has a large negative value.
For the nested case, $\mu=t'=0$, the insertion of
(\ref{RPA}) and (\ref{TMA}) into (\ref{SCRs2}) and (\ref{SCRd2})
lead to the followings;
\begin{eqnarray}
\xi_\sigma^{-2}&\!=&\!-(\sGamma-\frac{\usd}{\udd}\dGamma)/t\log^2\frac{E_c}{T}
+\xi_d^{-2}\usd/\udd
+(\uss-\frac{\usd^2}{\udd})\!
\int_0^{2E_c}\!\!\frac{d\omega}{\pi}\smint\!\frac{d^2{\mib k}}{(2\pi)^2}
{\rm Im}\chi_\sigma(\omega,{\mib k})\coth\frac{\omega}{2T}
\label{SCRs3} \\
\xi_d^{-2}&\!=&\!-(\dGamma-\frac{\usd}{\uss}\sGamma)/t\log^2\frac{E_c}{T}
+\xi_\sigma^{-2}\usd/\uss
+(\udd-\frac{\usd^2}{\uss})\!
\int_0^{2E_c}\!\!\frac{d\omega}{\pi}\smint\!\frac{d^2{\mib k}}{(2\pi)^2}
{\rm Im}\chi_d(\omega,{\mib k})\coth\frac{\omega}{2T}.
\label{SCRd3}
\end{eqnarray}
at sufficiently low temperatures.
Therefore ground state properties are determined by
$\sGamma-\dGamma\usd/\udd$ and $\dGamma-\sGamma\usd/\uss$.
If the former is positive and the latter is negative,
the system has the AFM order. If the former is negative and
the latter is positive, it has the $d$SC order.
If both values are positive, both orders may coexist
and thus $\pi$-triplet fluctuations are also required to consider.
If both values are negative, it belongs to the quantum-disordered region.


\begin{thebibliography}{99}
%
\bibitem{RMP}For a recent review see M. Imada, A. Fujimori and Y. Tokura: Rev. Mod. Phys. {\bf 70} (1998) 1039, Sec IV.C. 
%
%
\bibitem{LoeserShenDessauMarshallParkFournierKapitulnik1996}
Z. -X. Shen and D. S. Dessau: Physics Reports {\bf 253} (1995) 1;
A. G. Loesser {\it et al.}:
Science {\bf 273} (1996) 325.
%
%
%
\bibitem{photoemission}
%
%
%
%
H. Ding {\it et al.}:
Nature {\bf 382} (1996) 51;
%
%
%
%
%
%
D. S. Marshall {\it et al.}: Phys. Rev. Lett. {\bf 76} (1996) 4841.
%
%
\bibitem{Imada98}M. Imada, F. F. Assaad, H. Tsunetsugu and Y. Motome: cond-mat/9808044 and to be published.
\bibitem{Assaad98}F. F. Assaad and M. Imada: cond-mat/9811384.
\bibitem{RMPsecIIF11}For a recent review see M. Imada, A. Fujimori and
Y. Tokura: Rev. Mod. Phys. {\bf 70} (1998) 1039, Sec II.F.11.
%
%
%

\bibitem{Yasuoka}H. Yasuoka, T. Imai and T. Shimizu: ``Strong
Correlation and Superconductivity" ed. by H. Fukuyama, S. Maekawa and
A. P. Malozemoff (Springer Verlag, Berlin, 1989), p.254.

%
\bibitem{Y124NMR}H. Zimmermann, M. Mali, D.Brinkmann, J. Karpinski, E. Kaldis and S. Rusiecki:
Physica C {\bf 159} (1989) 681;
%
T. Machi, I. Tomeno, T. Miyataka, N. Koshizuka, S. Tanaka, T. Imai and H. Yasuoka:
Physica C {\bf 173} (1991) 32.
%
\bibitem{Bi2212NMR}K. Ishida, Y. Kitaoka, K. Asayama, K. Kadowaki and T. Mochiku:
Physica C {\bf 263} (1996) 371.
%
\bibitem{Hggap}Y. Itoh, T. Machi, A. Fukuoka, K. Tanabe, and
H. Yasuoka: J. Phys. Soc. Jpn. {\bf 65} (1996) 3751.  
%
\bibitem{Hg1223NMR}M. -H. Julien {\it et al.}:
Phys. Rev. Lett. {\bf 76} (1996) 4238.
%

\bibitem{Fong}H. F. Fong, B. Keimer, D. L. Milius and I. A. Aksay:
Phys. Rev. Lett. {\bf 78} (1997) 713.
\bibitem{RMP2} See Ref.~\cite{RMP} Sec.II.D,E,F and IV.C.
%
%
%
%
%
%
%

\bibitem{Leggett80}A. J. Leggett: J. Phys. (France) {\bf 41} (1980) C7.
\bibitem{NozieresSchmitt-Rink85}P. Nozi\`{e}re and S. Schmitt-Rink:
J. Low. Temp. Phys. {\bf 59} (1985) 195.
\bibitem{EmeryKivelson95}V. J. Emery and S. A. Kivelson:
Nature {\bf 374} (1995) 434.
\bibitem{pseudo-gap} M. Randeria: {\it cond-mat/9710223}, and {\it references therein}.
%
\bibitem{Hetrog9808051}B. C. den Hetrog: {\it cond-mat/9808051}.

\bibitem{Schulz87}H. J. Schulz: Europhys. Lett. {\bf 4} (1987) 609.
\bibitem{Dzyaloshinskii96}I. E. Dzyaloshinskii: J. Phys. I (France) {\bf 6}
(1996) 119.
\bibitem{FurukawaRiceSalmhofer9806159}N. Furukawa, T. M. Rice
and M. Salmhofer: Phys. Rev. Lett. {\bf 81} (1998) 3195.
%
%
\bibitem{SCR}
T. Moriya: {\it Spin Fluctuation in Itenerant Electron Magnetism} (Springer-Verlag, Berlin, 1985).
%
%
\bibitem{ChakravartyHalperinNelson89}S. Chakravarty, B. I. Halperin
and D. R. Nelson: Phys. Rev. B {\bf 39} (1989) 2344.
\bibitem{ChubukovSachdevYe94}A. V. Chubukov, S. Sachdev and J. Ye: Phys. Rev. B {\bf 49} (1994) 11919.
\bibitem{HertzKlenin74}J. A. Hertz and M. A. Klenin:
Phys. rev. B {\bf 10} (1974) 1084.
\bibitem{Hertz76}J. A. Hertz: Phys. Rev. B {\bf 14} (1976) 1165.
\bibitem{Millis93}A. J. Millis: Phys. Rev. B {\bf 48} (1993) 7183.
%
\bibitem{Gofron}K. Gofron, J. C. Campuzano, A. A. Abrikosov,
M. Lindroos, A. Bansil, H. Ding, D. Koelling and B. Dabrowski:
Phys. Rev. Lett. {\bf 73} (1994) 3302.
%
\bibitem{SachdevChubukovSokol95}S. Sachdev, A. V. Chubukov and A. Sokol:
Phys. Rev. B {\bf 51} (1995) 14874.
\bibitem{Schmalian9810041}J. Schmalian: {\it cond-mat/9810041}.
%
%
\bibitem{Zhang97science}S.-C. Zhang: Science {\bf 275} (1997) 1089.
\bibitem{Zhang98preprint}S.-C. Zhang: {\it cond-mat/9808309}.
%
%

\bibitem{K-Ttransition}L. Berezinski: Sov. Phys. JETP {\bf 32} (1970) 493;
J. M. Kostelitz and D. J. Thouless: J. Phys. C{\bf 6} (1973) 1181.
%
\bibitem{RanderiaVarlamov94}M. Randeria and A. Varlamov:
Phys. Rev. B {\bf 50} (1994) 10401.
%
\bibitem{MerminWagner66}M. D. Mermin and H. Wagner:
Phys. Rev. Lett. {\bf17} (1966) 1133.
%
\bibitem{MurakamiFukuyama98}M. Murakami and H. Fukuyama:
J. Phys. Soc. Jpn. {\bf 67} (1998) 2784.
%

%

%
%
%
%
\bibitem{MassiddaYuFreeman87-1}S. Massidda, J. Yu and A. J. Freeman:
Phys. Lett. A {\bf 122} (1987) 198.
%
\bibitem{SiZhaLevinLu93}Q. Si, Y. Zha, K. Levin and J. P. Lu:
Phys. Rev. B {\bf 47} (1993) 9055.
%
\bibitem{YBCO1236.9ARPES}J. C. Compuzano,
{\it et al.}:
Phys. Rev. Lett. {\bf 64} (1990) 2308.
%
\bibitem{MonthouxPines94}P. Monthoux and D. Pines:
Phys. Rev. B {\bf 50} (1994) 16015.
%
\bibitem{TsunetsuguImadaU}H. Tsunetsugu and M. Imada: unpublished.
\bibitem{Hg1201NMR}Y. Itoh, T. Machi, A. Fukuoka, K. Tanabe and
H. Yasuoka: J. Phys. Soc. Jpn. {\bf 65} (1996) 3751.
%
\bibitem{LSCONMR}T. Imai {\it et al.}:
J. Phys. Soc. Jpn. {\bf 59} (1990) 3846.
%
\bibitem{LSCOARPES}A. Ino {\it et al.}: {\it cond-mat/9809311};
Phys. Rev. Lett. {\bf 79} (1997) 2101;
Phys. Rev. Lett. {\bf 80} (1998) 2124.
%
%
\bibitem{Assaad96}F. F. Assaad, M. Imada, and D. J. Scalapino:
Phys. Rev. Lett. {\bf 77} (1996) 4592.
\bibitem{Assaad97}F. F. Assaad, M. Imada, and D. J. Scalapino:
Phys. Rev. B {\bf 56} (1997) 15001.
\bibitem{Assaad97b}F. F. Assaad, and M. Imada:
Phys. Rev. B {\bf 58} (1998) 1845.
%
\bibitem{Y1236.69neutron}J. Rossat-Mignod {\it et al.}:
Physica C {\bf 185-189} (1991) 86.
%
%
%
\bibitem{RGtheory}S. Onoda and M. Imada: unpublished.
%

\end{thebibliography}
\end{document}